\documentclass[a4paper,11pt]{article}
\usepackage{pos}
\usepackage{float}

\title{Taking dimuon production in DIS to NNLO precision}

\author{I. Helenius}
\author{H. Paukkunen}
\author*{S. Yrjänheikki}

\affiliation{University of Jyväskylä, Department of Physics, P.O. Box 35, 40014 University of Jyväskylä, Finland}

\affiliation{Helsinki Institute of Physics, P.O. Box 64, 00014 University of Helsinki, Finland}

\emailAdd{ilkka.m.helenius@jyu.fi}
\emailAdd{hannu.t.paukkunen@jyu.fi}
\emailAdd{sami.a.yrjanheikki@jyu.fi}

\abstract{We present an updated calculation of dimuon production in neutrino-nucleus collisions --- an important constraint of the relatively poorly-known strange-quark distribution --- at next-to-next-to-leading (NNLO) order. Compared to the approach usually used in the literature, where one computes inclusive charm production and multiplies the cross section by corrective factors, we instead compute the dimuon production cross section directly using semi-inclusive deep inelastic scattering (SIDIS) and a data-fitted decay function. While inclusive charm production has previously been computed to NNLO, we are using the recently-computed NNLO SIDIS coefficients to calculate the perturbative parts of the entire dimuon production process at NNLO. We find a notable reduction of scale uncertainties at larger momentum-fraction $x$, where the NNLO corrections tend to increase the cross section. At smaller $x$, we find a decrease in the cross section, a behaviour which can be linked to the suppression of the strange sea and which may eventually alleviate the tension observed between dimuon neutrino and LHC heavy-gauge boson data.}

\FullConference{33rd International Workshop on Deep Inelastic Scattering and Related Subjects (DIS2026)\\
 4-8 May 2026\\
Bologna, Italy\\}

\renewcommand{\hookAfterAbstract}{%
	\par\bigskip
	\textsc{ArXiv ePrint}:
	\href{https://arxiv.org/abs/2603.09646}{2603.09646}
}

\newcommand{\dd}{\mathrm{d}}

\begin{document}
\maketitle

\section{Introduction}

Global analyses of parton distribution functions (PDFs) rely on multiple datasets with different underlying partonic processes to constrain all parton flavors. One of the more challenging distributions to determine is that of the strange quark, which to this day is still poorly known \cite{Hou:2019efy,Bailey:2020ooq,NNPDF:2021njg,Eskola:2021nhw,Duwentaster:2022kpv,AbdulKhalek:2022fyi}. An important input to such a determination is dimuon production data in neutrino-nucleus collisions \cite{CCFR:1994ikl,NuTeV:2001dfo,NuTeV:2007uwm,CHORUS:2008vjb,NOMAD:2013hbk}. The underlying process one would like to utilize in this case is charm production. As the charm quark is not observable, one must instead look for the decay particles (a muon in this case) of the hadronized charm quark. While going from charm production to dimuon production is not trivial, it is typically treated in a simplified manner: the dimuon production cross section is obtained from charm production by simply multiplying it with corrective factors. Such a simplified approach works for a fixed-flavor leading-order calculation, but breaks down for higher orders or variable-flavor number schemes. Our framework, based on a semi-inclusive charmed-hadron production calculation and a decay function fitted to experimental data, improves on the simplified approach by calculating the full dimuon production process directly \cite{Helenius:2024fow,Paukkunen:2025kjb,Helenius:2026uuz}. In this work, we implement the NNLO corrections from ref. \cite{Bonino:2025qta} that have recently been calculated for SIDIS.

\section{The SIDIS framework for dimuon production}

In the SIDIS-based framework \cite{Helenius:2024fow,Paukkunen:2025kjb}, we write the full dimuon production cross section as
\begin{equation}
    \frac{\dd\sigma(\nu N\to \mu\mu X)}{\dd x \, \dd y}=\sum_h\int \dd z \frac{\dd\sigma(\nu N\to \mu hX)}{\dd x \, \dd y \, \dd z} B_{h\to\mu}(E_h=zyE_\nu, E_\mu^{\text{min}}),
\end{equation}
where the hadron-production process $\nu_\mu(k)+N(P_N)\to \mu(k')+h(P_h)+X$ is given by the usual SIDIS cross section
\begin{equation}
\label{eq:sidis_hadron_cross_section}
\begin{split}
	\frac{\dd\sigma(\nu_\mu N\to \mu h X)}{\dd x \, \dd y \, \dd z}= \frac{G_F^2M_W^4}{\left(Q^2+M_W^2\right)^2}
  \frac{Q^2}{2\pi xy}\bigg[&xy^2 F_1(x, z, Q^2)+\left(1-y-\frac{xy M^2}{s-M^2}\right)F_2(x, z, Q^2) \\ & \pm xy\left(1-\frac{y}{2}\right)F_3(x, z, Q^2)\bigg].
\end{split}
\end{equation}
Kinematics are described by the familiar invariants
\begin{equation}
\label{eq:sidis_kinematics}
\begin{aligned}
	Q^2&=-q^2 = -(k-k')^2, \quad x=\frac{Q^2}{2P_N\cdot q}, \quad y=\frac{P_N\cdot q}{P_N\cdot k}, \quad z=\frac{P_N\cdot P_h}{P_N\cdot q}.
\end{aligned}
\end{equation}
In the case of a (anti)neutrino beam, sign of the $F_3$ term is $+(-)$.

The purpose of the energy-dependent branching ratio $B_{h\to \mu}$ is to describe the decay of the charmed hadron $h$ into a muon. It depends on the hadron energy, and on an energy cut $E_\mu^{\text{min}}$ on the muon imposed by experiments. As it cannot be perturbatively computed, we parametrize it and then fit it to independent experimental data. More details on our framework can be found in refs.~\cite{Helenius:2024fow,Paukkunen:2025kjb,Helenius:2026uuz}.

\section{NNLO corrections}

Full charged-current NNLO SIDIS corrections have recently become available \cite{Bonino:2025qta,Goyal:2026ccx}. These corrections unlock new partonic channels that have previously been unavailable. Most relevant new channels are ones where a quark line connected to the PDF does not directly couple to the $W$ boson. For example, the $u\to c$ channel could potentially be significant due to the large valence contribution. 

In our previous work \cite{Paukkunen:2025kjb}, we have implemented the SIDIS process using the simplified Aivazis-Collins-Olness-Tung (SACOT) scheme with the rescaling variable $\chi=x(1+m_c^2/Q^2)$, known as the SACOT-$\chi$ scheme. Currently, the available NNLO corrections do not include mass corrections. Thus, we implement the NNLO calculation in an approximative SACOT-$\chi$ scheme, where the calculation to NLO uses the full SACOT-$\chi$ scheme, and the NNLO corrections are computed in a zero-mass scheme with the rescaling variable $\chi$. Including the rescaling variable takes the more relevant kinematic mass corrections into account, but this implementation still misses terms of the form $\mathcal{O}(\alpha_s^2 m_c^2/Q^2)$. 

\section{Results}

\begin{figure}[ht]
    \centering
    	\includegraphics[width=0.4\linewidth]{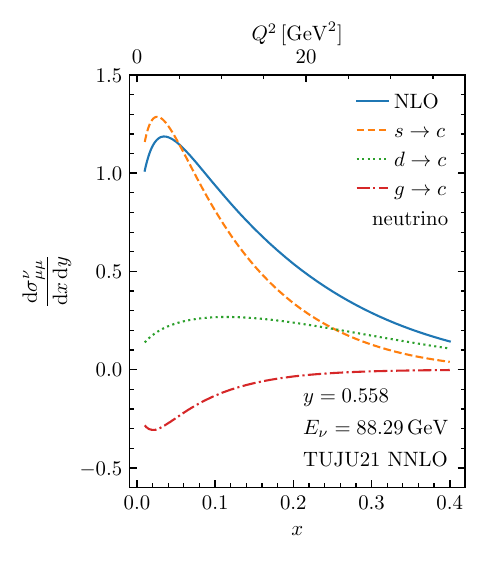}
        \includegraphics[width=0.4\linewidth]{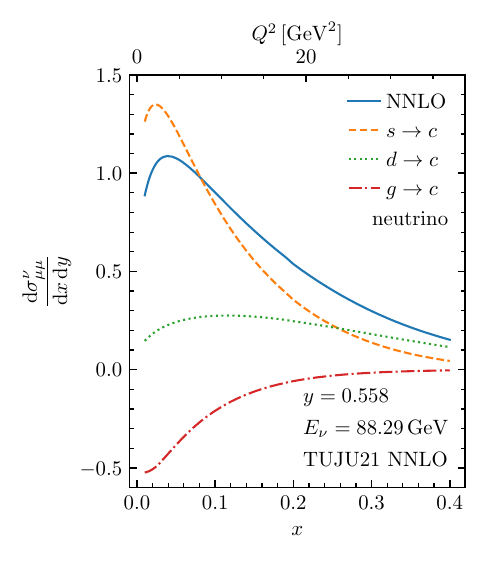}
    \caption{Channel decomposition of the dimuon production cross section at NLO (left) and NNLO (right).}
    \label{fig:full_nnlo_leading}
\end{figure}

\begin{figure}[ht]
    \centering
	\includegraphics[width=0.4\linewidth]{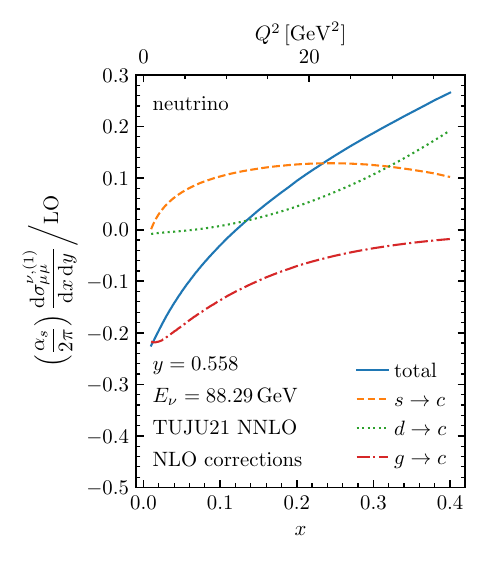}
    \includegraphics[width=0.4\linewidth]{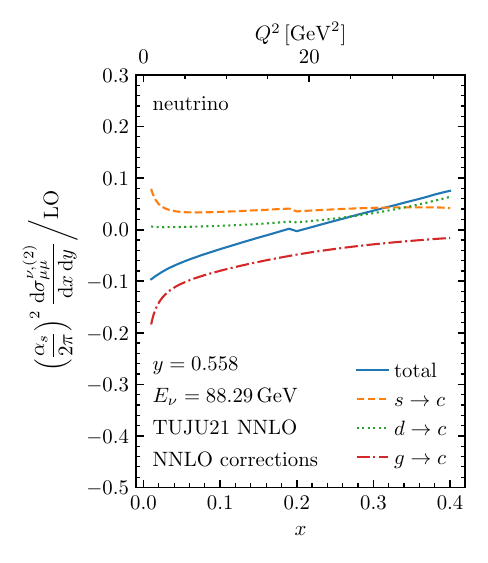}
    \caption{Leading-channel decomposition of the NLO (left) and NNLO (right) corrections, normalized to the LO cross section.}
    \label{fig:nnlo_leading}
\end{figure}

\begin{figure}[ht]
    \centering
    \includegraphics[width=0.4\linewidth]{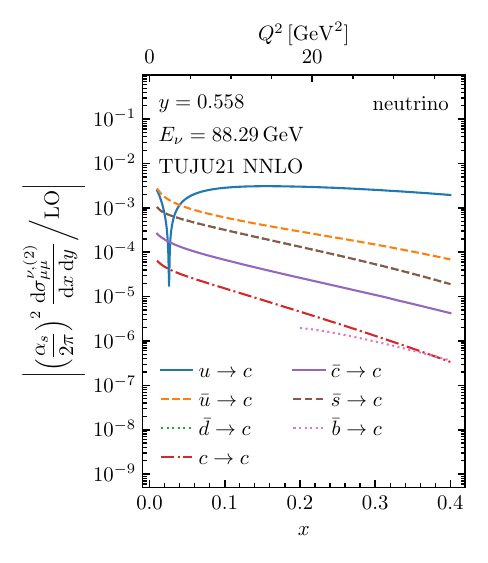}
    \includegraphics[width=0.4\linewidth]{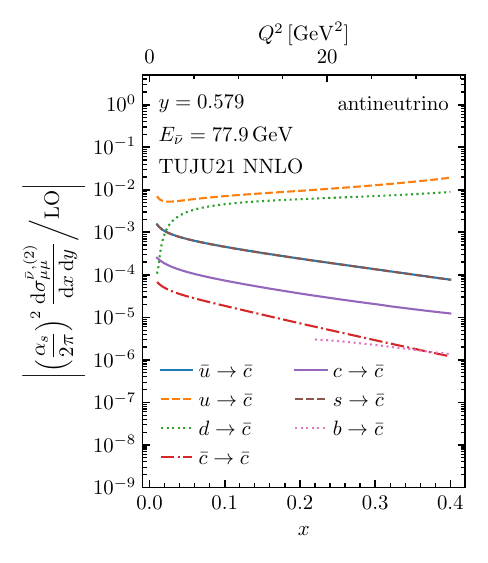}
    \caption{Subleading channel decomposition of the NLO (left) and NNLO (right) corrections, normalized to the LO cross section.}
    \label{fig:nnlo_subleading}
\end{figure}

Figure \ref{fig:full_nnlo_leading} shows how the new NNLO calculation impacts the leading channels of dimuon production. The cancellation between the $s$- and $g$-initiated channels grows at NNLO, i.e. the individual partonic channels are more impacted by the NNLO corrections than the cross section as a whole. The magnitudes of the NNLO corrections, shown in figure \ref{fig:nnlo_leading}, are smaller than the NLO corrections, indicating perturbative convergence. The gluon corrections remain similar, which is to be expected as they arise at NLO, making these NNLO corrections only the first corrections. The new partonic channels depicted in figure \ref{fig:nnlo_subleading} turn out to be insignificant, as they are orders of magnitude smaller than the corrections to the leading channels.

\begin{figure}[H]
    \centering
    \includegraphics[width=\textwidth]{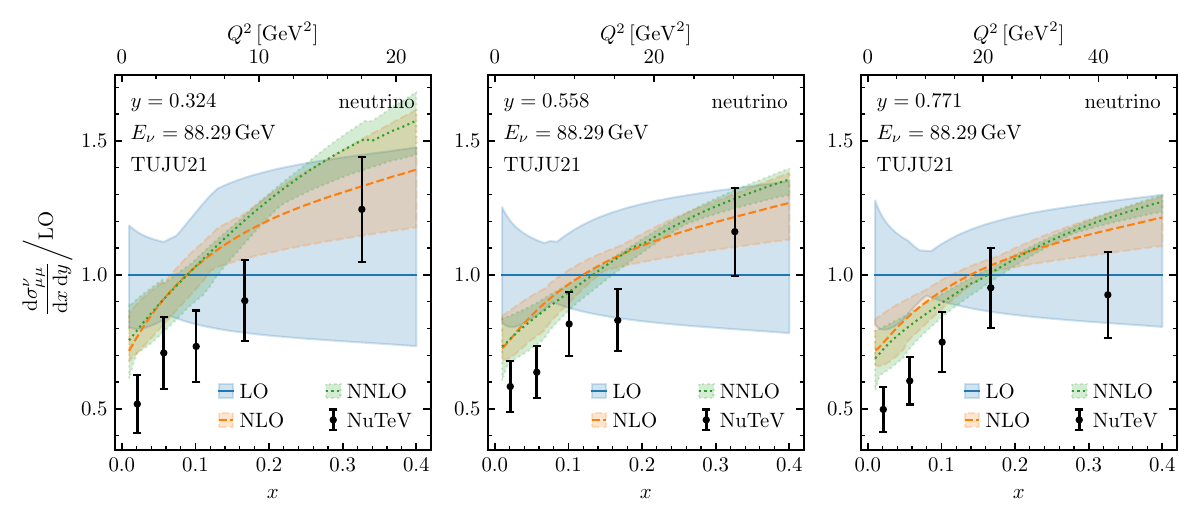}
    \caption{Scale uncertainties of the LO, NLO, and NNLO cross sections, normalized to the central LO value. Computed using the \texttt{TUJU21} PDF set.}
    \label{fig:nnlo_scales}
\end{figure}

Perturbative convergence is also indicated by the decrease in scale uncertainties at large $x$, which can be seen in figure \ref{fig:nnlo_scales}. The NLO and NNLO cross sections, however, coincide at smaller $x$, where the scale uncertainties are also the same. This behaviour could be due to the larger cancellation between the quark- and gluon-initiated channels; see figure \ref{fig:full_nnlo_leading}.

\begin{figure}[H]
    \centering
    	\includegraphics[width=\textwidth]{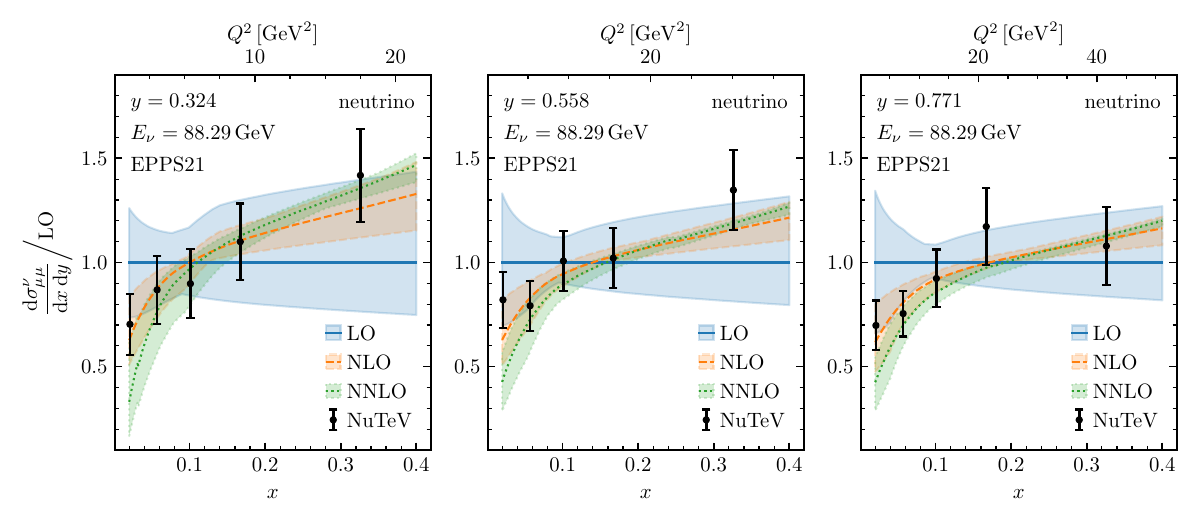}
    \caption{Same as figure \ref{fig:nnlo_scales}, but computed using the \texttt{EPPS21} PDF set.}
    \label{fig:epps}
\end{figure}

The strange-quark distribution can also be constrained using $W/Z$ production at the LHC. However, these collider data seem to be in tension with dimuon data, as collider data prefer an unsuppressed strange sea at $x\sim 0.02$ \cite{ATLAS:2012sjl,ATLAS:2016nqi} whereas neutrino data prefer a suppressed sea \cite{CCFR:1994ikl,NOMAD:2013hbk}. Figures \ref{fig:nnlo_scales} and \ref{fig:epps} show that the NNLO corrections alleviate this tension to some degree. The \texttt{TUJU21} PDF set (used in figure \ref{fig:nnlo_scales}) has an unsuppressed strange sea, whereas the central \texttt{EPPS21} set (used in figure \ref{fig:epps}) carries a suppressed sea. As was previously mentioned, the NLO and NNLO cross sections coincide at small $x$ in the case \texttt{TUJU21}. However, the NNLO corrections are more negative at small $x$ in the case of \texttt{EPPS21} due to the suppressed strange sea, which leads to a more pronounced role of the negative gluon-initiated channel. For both sets, the behaviour at larger $x$ is qualitatively the same. The difference in behaviour between the regions of larger and smaller $x$ alleviate the tensions between neutrino and collider data, since the preference of the collider data for an unsuppressed sea is specifically in the region of smaller $x$.

\section{Summary}

We have presented an update to our SIDIS-based framework for computing dimuon production in neutrino-nucleus collisions by taking the SIDIS calculation to NNLO precision. We have observed good perturbative convergence at larger $x$. We found that at smaller $x$, the behaviour of the NNLO corrections is tied to the suppression (or lack thereof) of the strange sea: the NLO and NNLO calculations coincide when using a PDF with an unsuppressed strange sea, whereas allowing for a suppressed sea results in the NNLO corrections being negative in this region. Such behaviour directly alleviates the tension between dimuon neutrino and collider data observed at $x\sim 0.02$.

\acknowledgments

This work has been supported by the Magnus Ehrnrooth foundation and the Center of Excellence in Quark Matter of the Research Council of Finland, project 364194 and by the Research Council of Finland project 361179. Finnish IT Center for Science, project jyy2580, is acknowledged for computing resources.

\bibliographystyle{JHEP}
\bibliography{Refs.bib}

\end{document}